# QCD in the chiral SU(3) limit from baryon masses on Lattice QCD ensembles

Matthias F.M. Lutz,[1] Yonggoo Heo,[2,1] and Renwick J. Hudspith[1]

[1]*GSI Helmholtzzentrum für Schwerionenforschung GmbH,*

*Planckstraße 1, 64291 Darmstadt, Germany*

[2]*Bogoliubov Laboratory for Theoretical Physics,*

*Joint Institute for Nuclear Research,*

*RU-141980 Dubna, Moscow region, Russia*

(Dated: October 10, 2024)

## Abstract

The baryon masses on CLS ensembles are used to determine the LEC that characterize QCD in the flavor-SU(3) limit with vanishing up, down, and strange quark masses. Here we reevaluate some of the baryon masses on flavor-symmetric ensembles with much-improved statistical precision, in particular for the decuplet states. These additional results then lead to a more significant chiral extrapolation of the Lattice data set to its chiral SU(3) limit. Our results are based on the chiral Lagrangian with baryon octet and decuplet fields considered at the one-loop level. Finite-box and discretization effects of the Lattice data are considered systematically. While in our global fit of the data we insist on large-$N_c$ sum rules for the LEC that enter at N$^3$LO, all other LEC are unconstrained. In particular, we obtain values for the chiral limit of the pion decay constant and the isospin-limit of the quark-mass ratio compatible with the FLAG report.

**CONTENTS**

## I. INTRODUCTION

Studies of QCD in its chiral limit are an important cornerstone of strong-interaction physics. While the steadily-improving Lattice QCD approach is most efficient at quark masses away from such a limit, the chiral Lagrangian is most reliable in the limit of vanishing up, down, and strange quark masses. The confluence of both approaches appears then to be the natural choice in establishing properties of QCD in its chiral limit.

The current challenge for the community is a quantitative control of Lattice QCD computations at small up, down, and small strange quark masses. This is complemented by the fact that the flavor-SU(3) Lagrangian, as used in conventional chiral perturbation theory, comes with rather poor convergence properties, in particular in the sector involving baryon fields. All together the determination of the low-energy constants (LEC) of the chiral Lagrangian, that characterize that chiral limit is a rather difficult problem. However, there is considerable progress at both ends: Lattice ensembles with small quark masses become more and more accessible as technology and techniques improve, but also the convergence properties of chiral SU(3) approaches appear much more favorable, if formulated in a resummed framework with on-shell hadron masses inside loop contributions. From the previous study [1] we expect reliable results for the baryon masses at quark masses as large as the physical strange quark mass.

In this work we will exclusively consider baryon masses as evaluated on Coordinate Lattice Simulations (CLS) ensembles. The Regensburg group (which is part of the CLS consortium) have published a large set of baryon octet and decuplet masses, mainly on CLS ensembles [2]. Since, in particular, the kaon masses on such ensembles are typically quite large, a quantitative reproduction of such data requires an analysis at N$^3$LO. This has recently been emphasized also in a dedicated study of the $\Omega$ baryon mass, which has been used to arrive at a precise determination of the lattice scale [3]. Any such analysis involves a large set of LEC, many of them are known only poorly at this stage. Nevertheless, some first results were obtained using relations amongst the set of LEC that arise in QCD in the limit of a large number of colors [1, 4]. While, in many cases such relations have turned quite useful, it is most desirable to abandon the use of such relations and allow the fit to the Lattice data to dictate what the LEC are.

Indeed, with the further-improved data set on the CLS ensembles relaxing these large-$N_C$

relations is more and more possible [3, 5–8]. As compared to our previous study [9] we will use values of the pion and kaon masses from [5], and $\Omega$ masses from [3]. In addition, we consider the more-recent determination of the baryon masses on the D200 ensemble from [10]. Moreover, a few updated values of the octet and decuplet masses along the $m = m_s$ trajectory have been performed specifically for this work. Most importantly, we do not constrain any of the LEC relevant at N$^2$LO by large-$N_c$ relations. Therefore our previous LEC may be subject to changes.

The work is organized as follows. In Section II we present more precise results on some flavor-symmetric CLS ensembles. Then we discuss in Section III how and which Lattice QCD data set is fitted. In Section IV we give a detailed presentation of our new fit scenario. It follows with Section V giving results on the various baryon sigma terms. Finally, we close with a summary and conclusions.

## II. BARYONS IN THE FLAVOR LIMIT

Any chiral extrapolation of baryon masses on some flavor-SU(3) Lattice QCD ensembles requires an accurate data set of the baryon octet and decuplet masses at degenerate quark masses. Owing to this need, the Regensburg group have supplemented the set of CLS ensembles by five additional flavor-symmetric ensembles, RQCD017, RQCD019, RQCD021, RQCD029 and RQCD030 [2]. Here we will scrutinize the rôle of these non-CLS ensembles with our fits. While there are also some flavor-symmetric CLS ensembles such as A650, A651, A652, A653, U103, H101, B450, H200, N202, N300, J500 or X250 available, such ensembles come typically with a pion mass above 300 MeV. Some ensembles with pion masses below this have been produced: namely X251, X450, and RQCD017. Of course, generating such ensembles from the lattice perspective is quite challenging, but for us these are of great importance. In order to monitor the path towards the chiral SU(3) limit, an accurate data set at smaller pion masses is highly desirable. Unfortunately, the Lattice data set of [2] shows quite limited precision along this mass trajectory, in particular for the decuplet masses. On all four low-pion-mass ensembles the decuplet masses are given with an uncertainty of about 10% only. Improved values for the baryon masses would be highly beneficial, and it is to this point that we now turn.

To remedy the low precision on these critical ensembles, we have undertaken efforts to improve the octet and decuplet mass determinations on the particular flavor-symmetric CLS ensembles: A650, A651, A652, X251, X450 and X250. Here we follow the approach of [3] by forming the generalized pencil of functions of appropriately parity-projected correlators

|            | A651 [2]    | current     | A652 [2]     | current     | A650 [2]     | current     |
|------------|-------------|-------------|--------------|-------------|--------------|-------------|
| $a\,m_\pi$ | 0.2751(9)   | 0.2748(7)   | 0.2140(10)   | 0.2140(8)   | 0.1835(13)   | 0.1829(13)  |
| $a\,M_N$   | 0.6715(44)  | 0.6763(18)  | 0.5842(41)   | 0.5837(16)  | 0.5469(54)   | 0.5443(22)  |
| $a\,M_\Omega$ | 0.8033(63) | 0.7985(22) | 0.6890(130) | 0.7033(21)  | 0.6630(130)  | 0.6668(21)  |

TABLE I. Updated determination of various hadron masses on the flavor-symmetric line for the coarsest lattice-spacing ($\beta = 3.34$) ensembles. Here we provide a comparison to the results given in [2].

|        | X250 [2]      | current     | X251 [2]     | current     | X450 [2]     | current     |
|--------|---------------|-------------|--------------|-------------|--------------|-------------|
| $a\,m_\pi$ | 0.1132(4)  | 0.1127(3)   | 0.0868(4)    | 0.0863(4)   | 0.1014(6)    | 0.1013(4)   |
| $a\,M_N$   | 0.3597(51) | 0.3508(11)  | 0.3185(85)   | 0.3197(10)  | 0.3764(61)   | 0.3688(7)   |
| $a\,M_\Omega$ | 0.4350(280) | 0.4331(11) | 0.3820(420) | 0.4042(9)  | 0.4902(68)   | 0.4683(8)   |

TABLE II. Updated determination of various hadron masses on the flavor-symmetric line for ensembles with light pions on finer ensembles. Here we provide a comparison to the results given in [2].

from simple Nucleon- or $\Omega$-like operators:

$$O_N = \epsilon_{abc}(u_a^T C\gamma_5 d_b)u_c, \qquad O_\Omega = \epsilon_{abc}(s_a^T C\gamma_i s_b)s_c. \tag{1}$$

These correlators are built from Coulomb gauge-fixed wall source propagators to greatly improve the statistical resolution. In addition, we utilize the Truncated Solver Method [11] to ameliorate further the cost of the calculation. Here we perform low-precision solves on every timeslice as the ensembles considered here are (anti-)periodic in time, and two high-precision solves (separated by $L_t/2$ with a randomly-chosen initial timeslice) per configuration. Typically we will use on the order of 16,000 spin and color diluted propagators per ensemble for the finer configurations and more than 40,000 for the smaller, coarse boxes.

Our results for the baryon masses are compiled into Tab. I and Tab. II and compared with previous results from [2]. We observe significant differences and emphasize that our values are statistically much more precise (particularly for the ensembles with a larger extent). The source of such differences is possibly explained by Fig. 1 where the effective mass for the octet and decuplet baryon masses on the X251 ensemble are shown using a single exponential ansatz for the correlation functions. We observe a much flatter effective mass using our gauge-fixed wall sources, as compared to the distinct choice of source functions in [2]. In Tab. III we collect the p-values and fit ranges of our exponential fits for the 6 ensembles discussed here. In all cases the single exponential ansatz implies a statistically sound reproduction of the effective masses. We note that in [3] a table corresponding to Tab. III is composed for further CLS ensembles applied in our work.

We also re-compute the pion mass on these ensembles using a slightly different method; stochastic $Z_2$ wall sources (PS-Wall in the parlance of [12]), on the same gauge configurations as the baryons were computed. These determinations along with the baryon masses are presented in Tabs. I. and II. Our pion masses agree with those of [2]. For further chiral fits to the full baryon spectrum we will use the $\Omega$-baryon determinations from [3] where available.

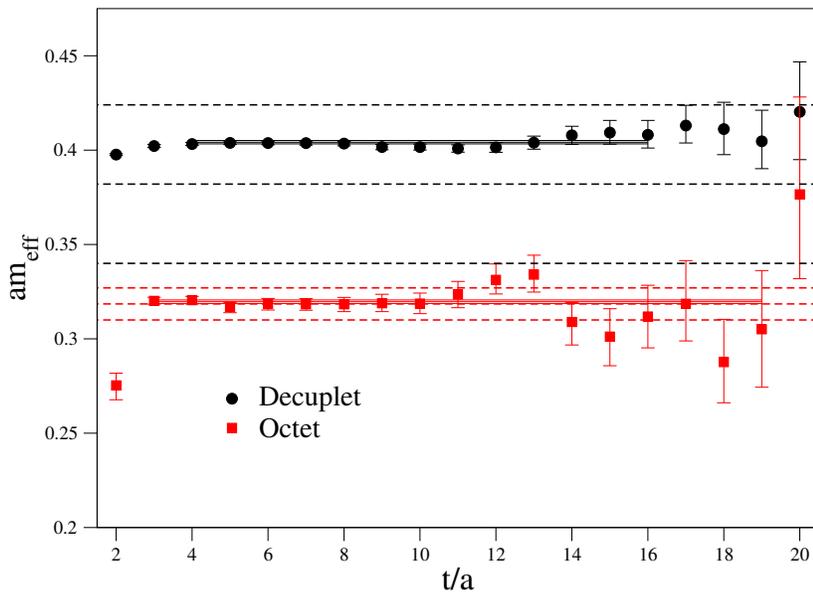

FIG. 1. Effective mass plots of the baryons on the X251 ensemble.

|   | $p_{A651}$ | range (t/a) | $p_{A652}$ | range (t/a) | $p_{A650}$ | range (t/a) |
|---|---|---|---|---|---|---|
| $N$ | 0.49 | 4-19 | 0.81 | 2-17 | 0.31 | 3-12 |
| $\Omega$ | 0.56 | 2-8 | 0.81 | 4-14 | 0.11 | 3-13 |
|   | $p_{X250}$ | range (t/a) | $p_{X251}$ | range (t/a) | $p_{X450}$ | range (t/a) |
| $N$ | 0.44 | 3-18 | 0.31 | 3-19 | 0.57 | 4-20 |
| $\Omega$ | 0.51 | 3-15 | 0.25 | 4-16 | 0.71 | 2-19 |

TABLE III. Fit ranges, together with their p-values of a single exponential ansatz for the effective mass functions on A651, A652, A650 and X250, X251, X450.

## III. BARYON MASSES ON CLS ENSEMBLES

We use the Lattice QCD results for meson and baryon masses on various CLS ensembles as made available by the Regensburg group in [2], unless re-computed by us. For given set of ensembles at fixed value of $\beta$ we determine an associated lattice scale, $a$, by enforcing that in the continuum limit at physical quark masses and infinite box size the isospin-averaged baryon octet and decuplet masses are reproduced.

Even though the ensembles used here are O(a)-improved via the nonperturbative clover term, at nonzero quark mass the lattice-spacing itself picks up an O(a)-mass dependent term [13, 14] from the gauge action, this then induces such a term in the conversion to physical quantities. This is of particular importance when combining trajectories that do not lie on the fixed Tr[M] trajectory. A way to effectively cancel these terms [2] is to perform a correction directly to the lattice-measured bare masses $M_{\text{hadron}}$,

$$a\, M_{\text{hadron}} \left(1 - a\, \bar{m}\, b_a\right), \qquad b_a = 0.31583(5) + \mathcal{O}(1/\beta), \qquad (2)$$

explicitly eliminating these O(a) effects. This is done with values for $a\,\bar{m}$ converted from $\kappa_l$ and $\kappa_s$ and $\kappa_{\text{crit}}(int)$ listed in [2] for the various ensembles. According to [2] the considered combination (2) receives further corrections from discretization effects at least quadratic in the lattice scale $a$ only. The parameter $b_a$ is so far only known at the one-loop level [2], and is related to the parameter $b_g$ where nonperturbative estimates have been attempted [15], albeit on finer lattices than used here. Therefore, we consider the value for $b_a$ as a free parameter with the expectation that it is likely large and positive.

In addition we permit the leading order LEC to depend on the lattice scale at $O(a^2)$. Following our previous work we write

$$M_{[8]} \to M_{[8]} + a^2\, \gamma_{M_8}, \qquad b_0 \to b_0 + a^2\, \gamma_{b_0}, \qquad b_D \to b_F + a^2\, \gamma_{b_D}, \qquad b_F \to b_F + a^2\, \gamma_{b_F},$$
$$M_{[10]} \to M_{[10]} + a^2\, \gamma_{M_{10}}, \qquad d_0 \to d_0 + a^2\, \gamma_{d_0}, \qquad d_D \to d_D + a^2\, \gamma_{d_D}, \qquad (3)$$

where such effects are implied by the framework developed in [16–19] for a Wilson quark action.

Our computation uses the pion and kaon masses as given in lattice units to determine the quark masses, $m$ and $m_s$, together with the eta mass by solving the finite-box variant of set of nonlinear equations (41) of [1]. This involves two parameter combinations, $2\,L_6 - L_4$ and $2\,L_8 - L_5$, where we recall that $3\,L_7 + L_8$ is determined by the request that the empirical eta mass is reproduced at physical values of the quark masses. With this, we can compute the baryon masses in terms of the finite-box variant of the set of nonlinear equations (7) of [9]. This involves a series of LEC, with their specific form recalled in the Appendix here for completeness. In this work we consider Lattice QCD data points with pion and kaon masses equal to or smaller than the empirical eta mass only. This is the most conservative assumption in a three-flavor extrapolation framework. The uncertainties in the lattice pion and kaon input masses are propagated in our fit by assuming a Gaussian distribution around their central values and widths determined by their given one-sigma errors.

Our error estimate of the LEC considers the 1-sigma statistical uncertainties of our fit, but also some systematic uncertainties as is implied by the limited accuracy of our chiral extrapolation framework. At this stage we assume a universal residual error that is estimated by the restriction that our Lattice data description reaches a $\chi^2/\text{dof} \simeq 1$. This is well-justified for ensembles with similar-sized value of $\bar{m}$, since the distance of the quark masses from their chiral-limit value should determine the accuracy at which the implied baryon masses can be computed. Our current study leads to a residual uncertainty of about 7 MeV, which is a significant reduction in systematic error in comparison to [9] where 12-14 MeV was deemed necessary.

## IV. A GLOBAL FIT

We report on our updated global fit of the baryon masses on CLS ensembles. A complete list of considered ensembles is given in Tab. IV, where we discuss in some detail our data choice. In most cases we use the values as published in [2] by the Regensburg group. Where available we improve the data set by using pion and kaon masses from [5]. Such ensembles are marked by an asterisk. Similarly we also assume the omega baryon masses from [3], as indicated by an additional row in the table. Not too surprsingly, our fit is based also on the improved baryon masses of Tab. I and Tab. II.

Some further ensembles generated by CLS are not taken into account. Typically, because a complete set of baryon masses is not available so far, or pion and kaon masses are outside our applicability domain. Also, the RQCD019 and RQCD029 ensembles come with pion masses above 600 MeV and therefore are not relevant in our study. In our fits we keep the

|            | $am_\pi$    | $am_K$      | $aM_B$  |         | $am_\pi$    | $am_K$      | $aM_B$  |
|------------|-------------|-------------|---------|---------|-------------|-------------|---------|
|            | $\beta=3.34$ |            |         | B451    | 0.1614(6)   | 0.2205(3)   | [2]     |
| A654*      | 0.1665(12)  | 0.2271(9)   | [3]     | X450    | 0.1013(4)   | 0.1013(4)   | Tab. II |
| A650       | 0.1829(13)  | 0.1829(13)  | Tab. I  | B450*   | 0.1606(5)   | 0.1606(5)   | [3]     |
| A653*      | 0.2119(9)   | 0.2119(9)   | [3]     |         | $\beta=3.55$ |            |         |
| A652       | 0.2140(8)   | 0.2140(8)   | Tab. I  | E250*   | 0.0417(4)   | 0.1592(1)   | [3]     |
|            | $\beta=3.40$ |            |         | D201    | 0.0648(4)   | 0.1630(2)   | [2]     |
| S100       | 0.0929(31)  | 0.2055(6)   | [2]     | D200*   | 0.0653(3)   | 0.1560(2)   | [3][10] |
| D101       | 0.0958(11)  | 0.2057(5)   | [2]     | N200*   | 0.0923(3)   | 0.1507(2)   | [3]     |
| C101*      | 0.0957(7)   | 0.2058(3)   | [3]     | N201    | 0.0928(3)   | 0.1704(2)   | [2]     |
| C102       | 0.0965(8)   | 0.2178(4)   | [2]     | S201    | 0.0945(5)   | 0.1523(4)   | [2]     |
| H106       | 0.1182(20)  | 0.2247(7)   | [2]     | N203*   | 0.1125(2)   | 0.1440(2)   | [3]     |
| U101       | 0.1184(24)  | 0.2005(13)  | [2]     | N204    | 0.1143(3)   | 0.1773(3)   | [2]     |
| N101*      | 0.1215(6)   | 0.2016(3)   | [3]     | X251    | 0.0863(4)   | 0.0863(4)   | Tab. II |
| H105*      | 0.1214(12)  | 0.2023(6)   | [3]     | X250    | 0.1127(3)   | 0.1127(3)   | Tab. II |
| H102*      | 0.1540(7)   | 0.1914(6)   | [3]     | N202*   | 0.1342(3)   | 0.1342(3)   | [3]     |
| U102       | 0.1550(8)   | 0.1925(6)   | [2]     | H200*   | 0.1354(6)   | 0.1354(6)   | [3]     |
| H107       | 0.1592(7)   | 0.2375(5)   | [2]     |         | $\beta=3.70$ |            |         |
| RQCD021    | 0.1470(9)   | 0.1470(9)   | [2]     | E300*   | 0.0439(2)   | 0.1237(1)   | [3]     |
| U103 [20]  | 0.1803(6)   | 0.1803(6)   | [3]     | J303*   | 0.0648(2)   | 0.1196(2)   | [3]     |
| H101*      | 0.1822(6)   | 0.1822(6)   | [3]     | J304    | 0.0655(2)   | 0.1318(1)   | [2]     |
|            | $\beta=3.46$ |            |         | N302*   | 0.0869(3)   | 0.1136(3)   | [3]     |
| D450*      | 0.0833(4)   | 0.1838(2)   | [3]     | N304    | 0.0886(3)   | 0.1396(3)   | [2]     |
| D451       | 0.0837(3)   | 0.1939(2)   | [2]     | N305    | 0.1072(4)   | 0.1460(3)   | [2]     |
| N450       | 0.1097(3)   | 0.2018(2)   | [2]     | N300*   | 0.1057(2)   | 0.1057(2)   | [3]     |
| N401       | 0.1099(6)   | 0.1776(4)   | [2]     |         | $\beta=3.85$ |            |         |
| B452       | 0.1349(5)   | 0.2097(3)   | [2]     | J501*   | 0.0658(2)   | 0.0879(2)   | [3]     |
| S400*      | 0.1351(4)   | 0.1702(4)   | [3]     | J500*   | 0.0815(2)   | 0.0815(2)   | [3]     |

TABLE IV. List of CLS ensembles with their $a\,m_\pi$ and $a\,m_K$ values as taken in our work.

RQCD021 but do not consider the impact of RQCD017 and RQCD030, simply because our updated masses on the X250, X251, and X450 boxes are of much improved quality. The consideration of such results on the latter ensembles has a crucial impact on our fits since they depend decisively on ensembles with degenerate quark masses. Overall, an excellent reproduction of the data is achieved, with only some tension typically on the ensembles with the smallest pion masses. We emphasize that, contrary to our previous fits, the nucleon and lambda masses on the D450 and D451 ensembles can be recovered with reasonable accuracy. Also, the baryon masses on the ensembles E250 are considered in our current $\chi^2$, which has a pion mass of about 130 MeV - smaller than its empirical value. For this ensemble the isobar state is characterized most likely by more than one finite-box energy level. In this case, our conviction to use on-shell masses for the baryons inside loop contributions becomes ambiguous, and the framework would need an extension. Nevertheless, the tension built up by this challenge seems much reduced as compared to our previous fit. In order to avoid the fit to be pulled into uncontrolled territory we find we must exclude the $\Delta$ and $\Sigma^*$ on ensembles with $L > 4$ fm in our $\chi^2$ function; removing the D101, D450, D451, E250, D201, D200 and E300 ensembles, that all come with pion masses smaller than about 200 MeV. In addition, the ensemble D150 with the smallest pion masses of about 125 MeV is not included into our $\chi^2$ function at all. In the end, we are left with 348 data points to be approximated. We obtain a $\chi^2$ per degree of freedom of about 0.988 in our best fit, where we assumed a residual error of 7 MeV only.

|         | Fit [9]     | current Fit    |
|---------|-------------|----------------|
| $f$ [MeV] | 92.4*     | 82.35(68)      |
| $F_A$   | 0.51*       | 0.4852(73)     |
| $D_A$   | 0.72*       | 0.4855(85)     |
| $C_A$   | 1.44*       | 0.9740(415)    |
| $H_A$   | 2.43*       | 1.8390(188)    |
| $M$ [MeV] | 804.3(1)  | 840.3(15.7)    |
| $M + \Delta$ [MeV] | 1115.2(1) | 1091.2(13.8) |

TABLE V. The leading symmetry-preserving LEC from a fit to the baryon octet and decuplet masses. The values in the second to last column show the result of a previous fit to the Regensburg data set [2]. An asterisk is used if the LEC was not fitted.

In Tab. V we present the leading-order LEC that are relevant in the chiral limit. They are compared with values used in our previous fit of baryon masses on the CLS ensembles extended by some RQCD ensembles. A comparison of the two columns reveals a stunning effect in both of the baryon mass parameters, $M$ and $M + \Delta$. This is clearly related to the omission of the Regensburg ensembles, but also in part to the fact that we changed the fit strategy by not using large-$N_c$ relations for the axial coupling constants $F_A, D_A, C_A$ and $H_A$. We find it interesting, that for the first time such a fit of the baryon masses leads to a chiral limit value pion decay constant compatible with the value $f = 80.3(6.0)$ from FLAG [21]. We tried such fits in the former setup, but were never able to obtain a reasonable result compatible with such a value.

Here, we further scrutinize the consequence of such a scenario. In Tab. VIII we show the LEC that characterize the translation of pion and kaon masses to the quark masses. We find it striking, that for the first time our current fit suggests a quark mass ratio $m_s/m$ that is also quite close to the current FLAG report value 27.42(12) [21]. Like in our previous determinations we find values for the Gasser and Leutwyler LEC that are quite compatible with corresponding fits [1, 22] to the previous world Lattice data on the baryon masses [23–28], despite the fact that no attention was paid to possible discretization effects. Our values can be compared with dedicated studies not based on baryon masses from MILC, HPQCD or phenomenology [29–31]. Such values are much less precise as compared to our

|  |  | Fit [9] | current Fit |  |  | Fit [9] | current Fit |
|---|---|---|---|---|---|---|---|
| $b_0$ | [GeV$^{-1}$] | -0.8144(9) | -0.5941(795) | $d_0$ | [GeV$^{-1}$] | -0.4347(14) | -0.3723(674) |
| $b_D$ | [GeV$^{-1}$] | 0.1235(2) | 0.0529(111) | $d_D$ | [GeV$^{-1}$] | -0.5169(13) | -0.6938(174) |
| $b_F$ | [GeV$^{-1}$] | -0.2820(3) | -0.2931(134) | $10^3 (L_8 + 3 L_7)$ | | -0.4768(4) | -0.3145(28) |
| $10^3 (2 L_6 - L_4)$ | | 0.0411(3) | 0.0296(41) | $10^3 (2 L_8 - L_5)$ | | 0.0826(12) | -0.0769(51) |
| $m_s/m$ | | 26.15(1) | 27.62(4) | | | | |

TABLE VI. The leading symmetry-breaking LEC from a fit to the baryon octet and decuplet masses of the chiral Lagrangian. The values in the second last column show the result of a previous fit to the Regensburg data set [2].

|  | Fit [9] | current Fit |  | Fit [9] | current Fit |
|---|---|---|---|---|---|
| $a_{\text{CLS}}^{\beta=3.34}$ [fm] | 0.09337(22) | 0.09506(142) | $a_{\text{CLS}}^{\beta=3.55}$ [fm] | 0.06314(12) | 0.06344(23) |
| $a_{\text{CLS}}^{\beta=3.40}$ [fm] | 0.08251(7) | 0.08451(30) | $a_{\text{CLS}}^{\beta=3.70}$ [fm] | 0.05003(16) | 0.04963(4) |
| $a_{\text{CLS}}^{\beta=3.46}$ [fm] | 0.07478(8) | 0.07477(21) | $a_{\text{CLS}}^{\beta=3.85}$ [fm] | 0.03845(11) | 0.03798(7) |
| $\gamma_{M_8}$ [GeV$^3$] | -0.1322(10) | -0.2239(1221) | $\gamma_{M_{10}}$ [GeV$^3$] | -0.0776(4) | -0.2606(276) |
| $\gamma_{b_0}$ [GeV] | 0.0619(8) | -0.0014(789) | $\gamma_{d_0}$ [GeV] | -0.0115(9) | -0.0455(169) |
| $\gamma_{b_D}$ [GeV] | -0.1512(9) | -0.0779(187) | $\gamma_{d_D}$ [GeV] | 0.0206(9) | -0.0619(378) |
| $\gamma_{b_F}$ [GeV] | -0.0071(4) | -0.0303(70) | $b_a$ | 0.6305(8) | 0.7920(167) |

TABLE VII. Our determination of the lattice scales for the CLS ensembles, together with the various discretization parameters introduced in (2) and (3).

determination from the baryon masses. For instance:

$$10^3 \, (2\, L_6 - L_4)_{\text{MILC}} = 0.04(24)^{32}_{27}, \qquad 10^3 \, (2\, L_8 - L_5)_{\text{MILC}} = -0.20(11)^{45}_{19},$$
$$10^3 \, (2\, L_6 - L_4)_{\text{HPQCD}} = 0.23(17), \qquad 10^3 \, (2\, L_8 - L_5)_{\text{HPQCD}} = -0.15(20). \qquad (4)$$

We confirm from [2] that even at our improved-precision determination, the value for $2\, L_8 - L_5$ is close to zero at $\mu = 0.77$ GeV.

More details on the fit itinerary are collected in Tab. VII, where we present our result for the lattice scale parameters, together with the discretization parameters $\gamma_{...}$ that drive the size of discretization effects. It is worth mentioning that our lattice scales agree better and better with the values obtained in [2] as the $\beta$ value rises, implying a more and more continuum-like system, as they should. Like in all our previous analyses of Lattice QCD baryon masses, the quite precise determination of the lattice scales [1, 9, 22, 32] is characteristic of studies that use the on-shell hadron masses inside loop contributions to the baryon masses, but also perform a scale-setting in terms of the set of empirical baryon octet and decuplet masses. Despite the fact that we use a residual systematic error of 7 MeV in the baryon masses, the precision at which the lattice scales are determined appears competitive with the previous dedicated study of omega baryon masses on CLS ensembles [3].

Upon an inspection of (3) the impact of the $\gamma$'s on the LEC appears smaller than 5%

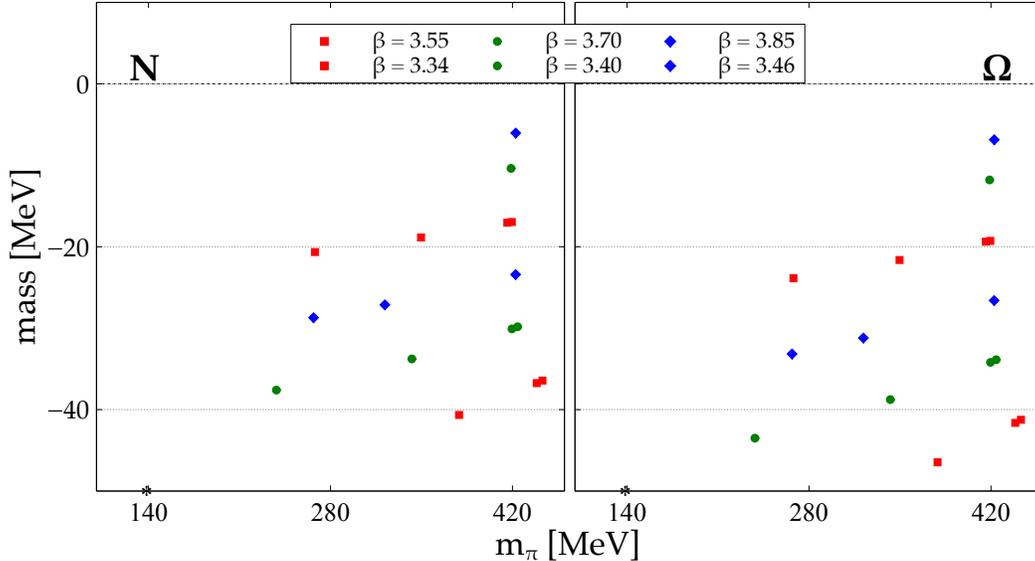

FIG. 2. Size of discretization effects in the baryon masses on flavor-symmetric ensembles.

even on the coarsest lattice with our smallest $\beta$ value. It is quite moderate in almost all cases. In Fig. 2 we show the total size of the discretization effects for the baryon octet and decuplet masses on flavor-symmetric ensembles. To illustrate such contributions we used the central values from Tab. VII.

We wish to comment on the possible addition of correlation effects that go beyond the correlations implied by our scale setting procedure. As we are using data from a wide range of sources for the pions and kaons as well as the baryon masses incorporating their correlations is not possible at this stage. Moreover, the residual systematic error that we enforce by hand decorrelates the data further by some degree by introducing something akin to uncorrelated noise on top of each measurement.

We complete our list of LEC with Tab. VIII and Tab. XI. The higher order LEC are grouped into symmetry-preserving in Tab. VIII versus symmetry-violating terms in Tab. XI, which we provide in the Appendix. All LEC are reasonably-well determined in our fit and have natural size. Their values differ in part significantly from our previous results in [1, 9, 22], which we attribute to the prior neglect of discretization effects, or a change in the fitted data sets. As compared to our previous analysis [9] the one-sigma error given for our current fit turns out significantly larger. This we trace back to the large-$N_c$ constraints used for the axial-coupling constants $F_A, D_A, C_A$ and $H_A$ in [9], which were not imposed in the current study.

| | Fit [9] | current Fit | | Fit [9] | current Fit |
|---|---|---|---|---|---|
| $g_0^{(S)}$ [GeV$^{-1}$] | -7.1454(14) | -6.2128(1.5691) | $g_0^{(V)}$ [GeV$^{-2}$] | 1.5983(7) | 5.0764(740) |
| $g_1^{(S)}$ [GeV$^{-1}$] | -0.6454(1) | -3.4537(3612) | $g_1^{(V)}$ [GeV$^{-2}$] | -4.1566(12) | -4.0551(576) |
| $g_D^{(S)}$ [GeV$^{-1}$] | -0.8087(14) | 0.9229(6918) | $g_D^{(V)}$ [GeV$^{-2}$] | 3.2166(10) | 6.1611(1525) |
| $g_F^{(S)}$ [GeV$^{-1}$] | -2.9798(9) | -3.8802(6657) | $g_F^{(V)}$ [GeV$^{-2}$] | -3.9400(3) | 1.3456(1208) |
| $h_1^{(S)}$ [GeV$^{-1}$] | -3.0444(2) | -3.3828(7795) | $h_1^{(V)}$ [GeV$^{-2}$] | 3.8897(1) | 4.1857(2760) |
| $h_2^{(S)}$ [GeV$^{-1}$] | 0. | 0. | $h_2^{(V)}$ [GeV$^{-2}$] | -0.4844(82) | 7.7346(2270) |
| $h_3^{(S)}$ [GeV$^{-1}$] | -1.9280(3) | -5.6282(5798) | $h_3^{(V)}$ [GeV$^{-2}$] | -4.8073(60) | -1.5500(765) |
| $h_4^{(S)}$ [GeV$^{-1}$] | -1.7500(2) | -6.3879(6177) | $h_5^{(S)}$ [GeV$^{-1}$] | -4.9384(10) | -4.0818(3811) |
| $h_6^{(S)}$ [GeV$^{-1}$] | 1.7500(7) | 6.3879(6177) | | | |

TABLE VIII. LEC from a fit to baryon masses on CLS ensembles. Owing to the large-$N_c$ sum rules as derived in [1, 4], there are 11 independent LEC here only.

The set of LEC in Tab. VIII characterize the scattering of Goldstone bosons off the baryon fields in the chiral limit at low energies. They are the basis of coupled-channel studies of such processes at physical quark masses, for which there exists already a large data basis from pion- and kaon-beam experiments.

Our results for the baryon octet masses on the various ensembles are shown in Fig. 3 and 4. We use our lattice scales in combination with our $b_a$ value as already specified in Tab. VII to convert to GeV. The Lattice data are shown in colored symbols always, where a given color corresponds to ensembles at a particular fixed-$\beta$ value. The chiral EFT results are in white symbols, where the shown error size is the 7 MeV estimate as discussed above. For all ensembles a good reproduction of the nucleon, lambda, sigma and xi masses is obtained. This is so even for the very-light D150 ensemble (the left-most data point) in Fig. 3, for which we chose to recall the Lattice values in yellow symbols, as to remind the reader that those masses were not included in our fit.

The distinct nature of the three group of ensembles: fixed sum of quark masses, fixed strange quark mass, and degenerate quark masses is clearly seen for all baryon masses with non-vanishing strange-quark content. The quite accurate lattice masses on the flavor-symmetric trajectories are the achievement of the current study, with updated high-precision baryon masses on the X250, X251 and X450 ensembles. The results on the RQCD017 and RQCD030 are to be recalled by yellow symbols in Fig. 3. Even given their rather large uncertainties, they are compatible with our fit result although they are not fitted directly.

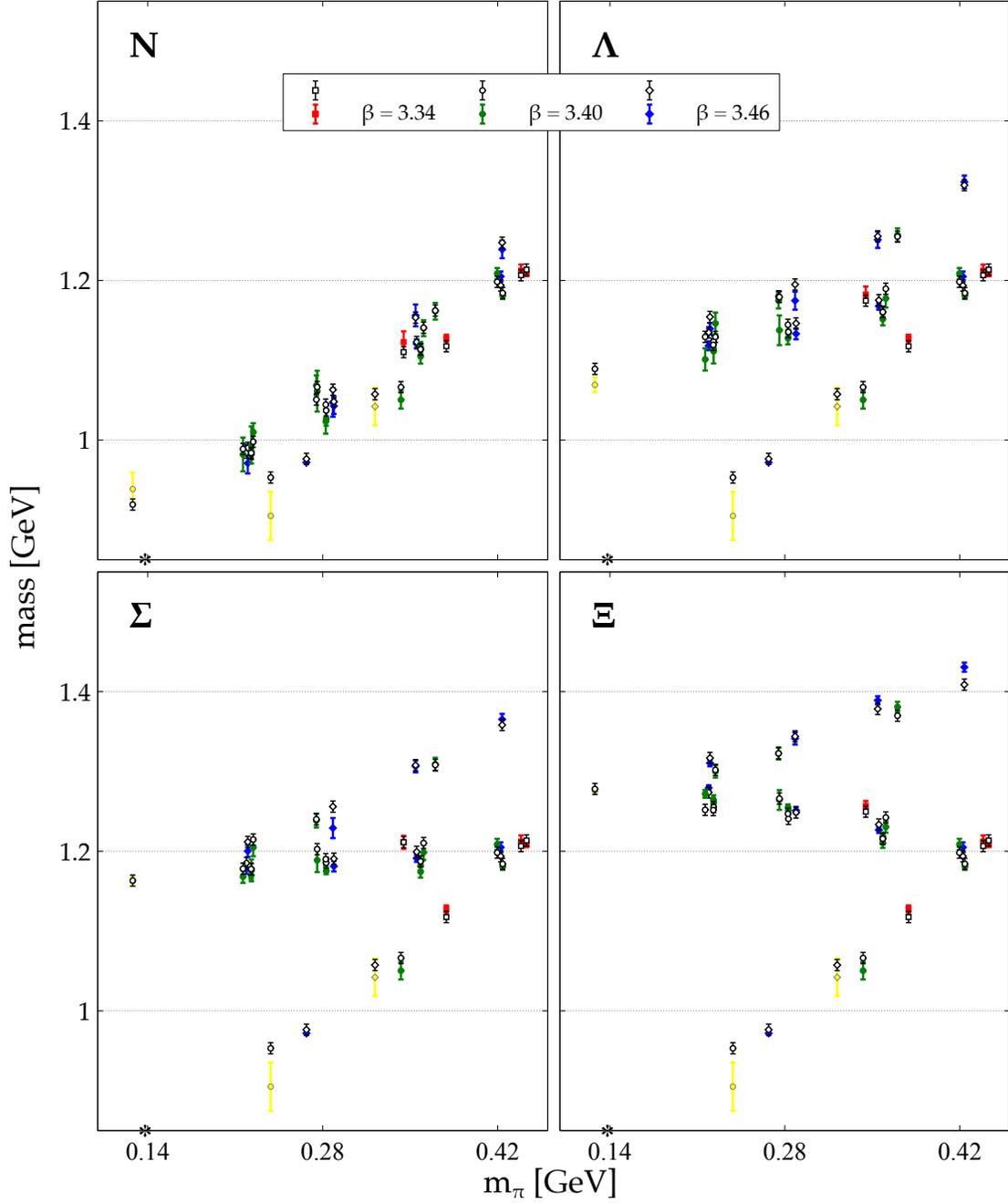

FIG. 3. Baryon octet masses on CLS ensembles as a function of the pion mass on the three coarsest $\beta$ values [2]. Lattice data that did not enter our $\chi^2$ function are shown in yellow.

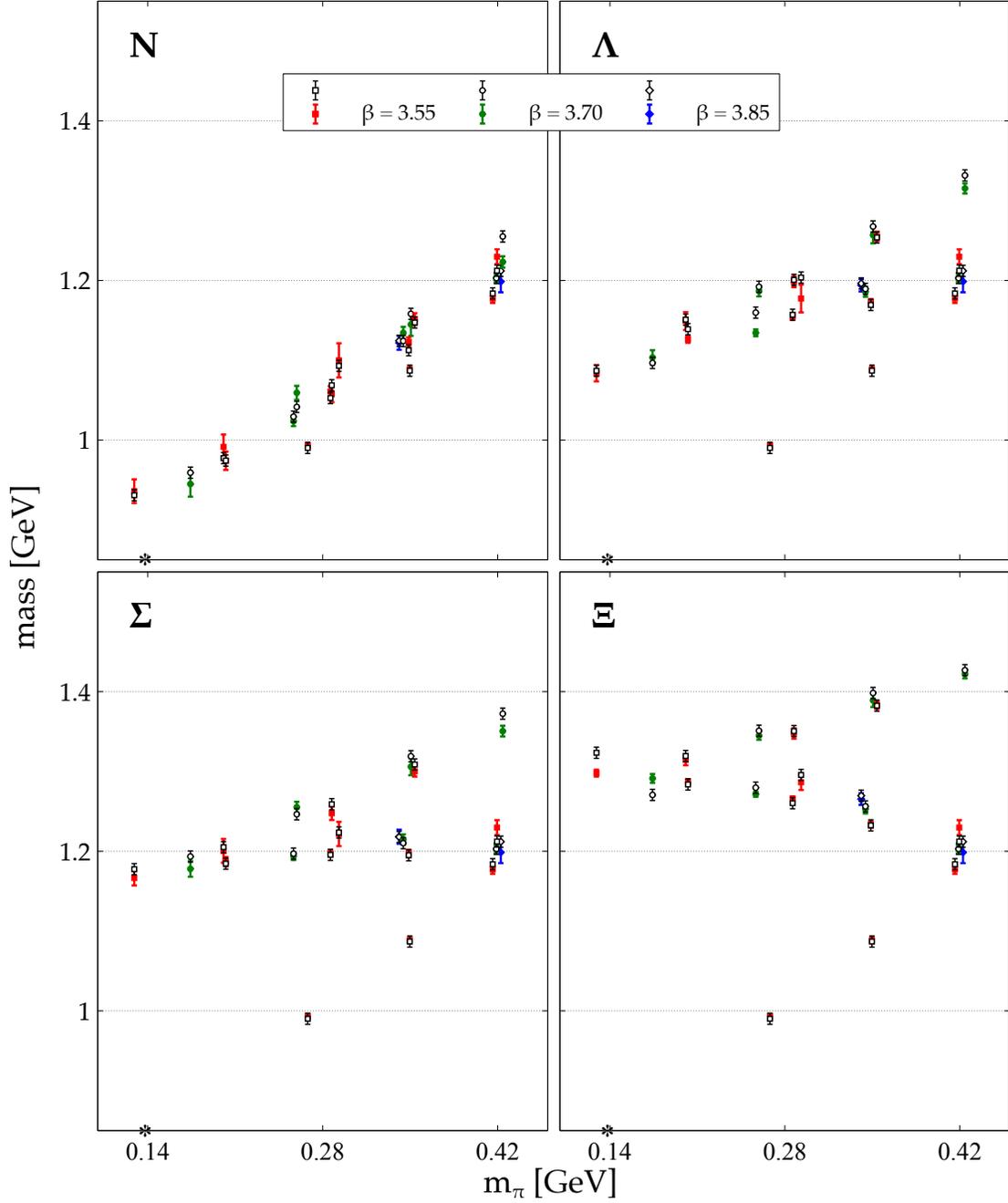

FIG. 4. Baryon octet masses on CLS ensembles as a function of the pion mass on the three finest $\beta$ values [2]. Lattice data that did not enter our $\chi^2$ function are shown in yellow.

It is clear that currently such data points appear not very well suited to determine the baryon masses along the flavor-symmetric line. Here the much more precise and novel results on the X251, X250 and X450 of Tab. II are instrumental.

Our results for the baryon decuplet masses on the various CLS ensembles are shown in

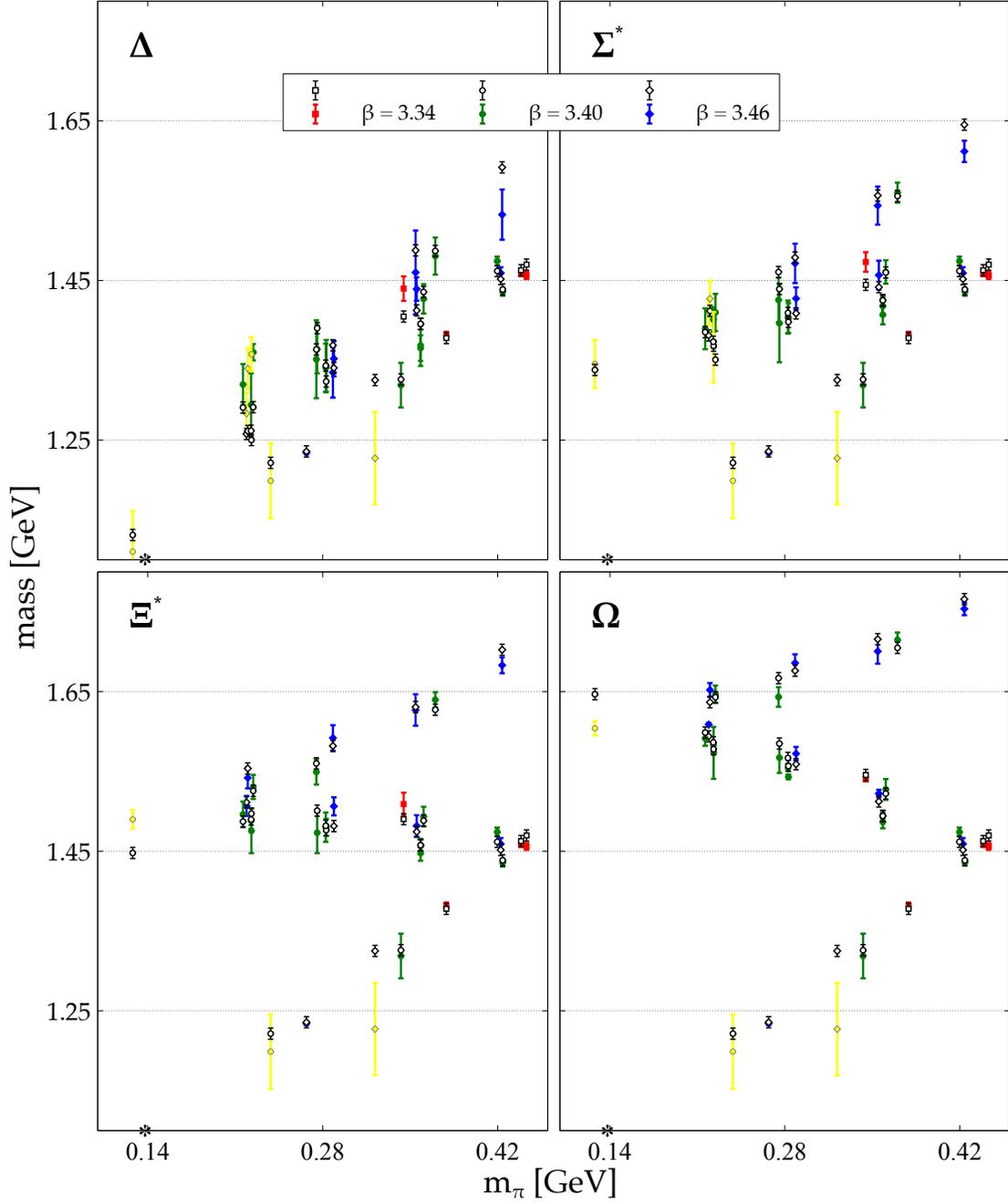

FIG. 5. Baryon decuplet masses on CLS ensembles as a function of the pion mass for the three coarsest $\beta$ values [2, 3]. Lattice data that did not enter our $\chi^2$ function are shown in yellow.

Fig. 5 and 6. Again, the results omitted from our fit for RQCD017 and RQCD030 are given by yellow symbols in Fig. 5. Additional yellow symbols also in Fig. 6 for the $\Delta$ and $\Sigma^*$ at small pion masses are to remind the reader that such results did not enter our $\chi^2$ function. We expect that our one-level approximation for such baryon resonances starts to become invalid around these pion masses, in particular the box size is tuned in an unfortunate manner such that $m_\pi L$ is kept large. Nevertheless, as we consider ensembles with larger pion masses or smaller box sizes we again obtain an accurate reproduction of the Lattice data set.

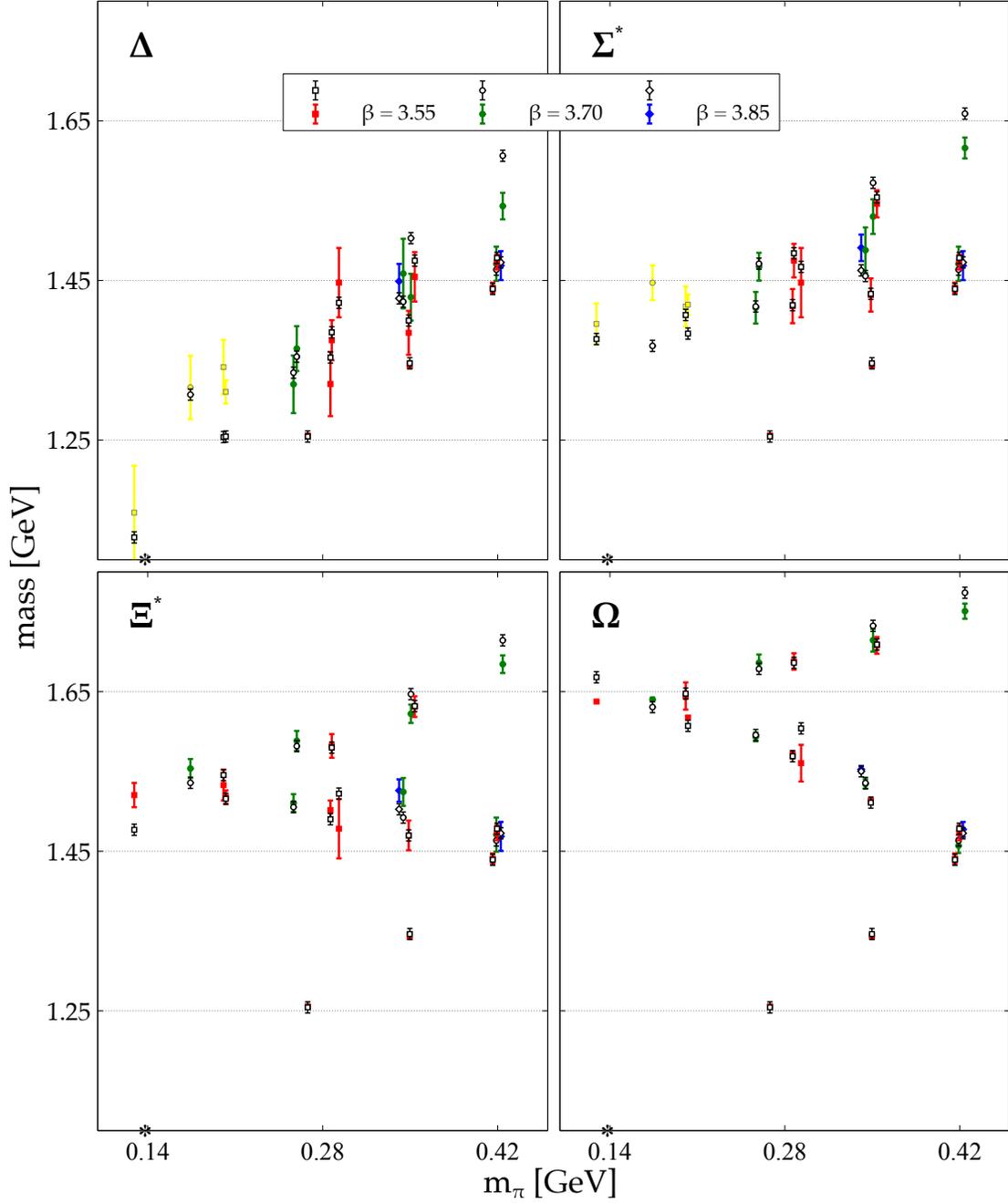

FIG. 6. Baryon decuplet masses on CLS ensembles as a function of the pion mass for the three finest $\beta$ values [2, 3]. Lattice data that did not enter our $\chi^2$ function are shown in yellow.

|  | [37] | [38] | [32] | [2] | current Fit |
|---|---|---|---|---|---|
| $\sigma_{\pi N}$ | $39(4)^{+18}_{-7}$ | $31(3)(4)$ | $39(2)$ | $43.9^{(4.7)}_{(4.7)}$ | $44.2(1.4)$ |
| $\sigma_{\pi\Lambda}$ | $29(3)^{+11}_{-5}$ | $24(3)(4)$ | $23.1(5)$ | $28.2^{(4.3)}_{(5.4)}$ | $40.3(9)$ |
| $\sigma_{\pi\Sigma}$ | $28(3)^{+19}_{-3}$ | $21(3)(3)$ | $18.3(5)$ | $25.9^{(3.8)}_{(6.1)}$ | $20.5(8)$ |
| $\sigma_{\pi\Xi}$ | $16(2)^{+8}_{-3}$ | $16(3)(4)$ | $5.7(1.0)$ | $11.2^{(4.5)}_{(6.4)}$ | $22.0(9)$ |
| $\sigma_{sN}$ | $34(14)^{+28}_{-24}$ | $71(34)(59)$ | $84(15)$ | $16^{(58)}_{(68)}$ | $-1(8)$ |
| $\sigma_{s\Lambda}$ | $90(13)^{+24}_{-38}$ | $247(34)(69)$ | $230(6)$ | $144^{(58)}_{(76)}$ | $112(10)$ |
| $\sigma_{s\Sigma}$ | $122(15)^{+25}_{-36}$ | $336(34)(69)$ | $355(5)$ | $229^{(65)}_{(70)}$ | $213(8)$ |
| $\sigma_{s\Xi}$ | $156(16)^{+36}_{-38}$ | $468(35)(59)$ | $368(13)$ | $311^{(72)}_{(83)}$ | $380(16)$ |

TABLE IX. Pion and strangeness sigma terms of the baryon octet states in units of MeV. A comparison with various other predictions [2, 32, 37, 38] is provided.

## V. BARYON SIGMA TERMS

The values of baryon sigma terms are a critical input to various topical challenges, such as in dark matter searches (see e.g. [33, 34]), a possible kaon condensate in dense nuclear matter [35], or a possible chiral symmetry restoration in dense nuclear matter (see e.g. [36]). The pion-baryon sigma terms, $\sigma_{\pi B}$, and the strangeness sigma terms, $\sigma_{sB}$, are introduced by the equations:

$$\sigma_{\pi B} = m\,\frac{\partial}{\partial m}M_B\,, \qquad \sigma_{sB} = m_s\,\frac{\partial}{\partial m_s}M_B\,, \qquad (5)$$

at physical quark masses, $m$ and $m_s$, and infinite volume in the continuum limit. We work here in the isospin-limit, where the averages of empirical values from the PDG are used. From the knowledge of the quark-mass dependence of the baryon masses such sigma terms can readily be determined. For given quark masses the physical baryon masses, as well as the pion, kaon, and eta masses are reproduced by construction. Small variations around such quark masses as requested by the derivative in (5) are followed up fully, by numerical solutions of our set of coupled and nonlinear equations that determine the meson and baryon masses in our framework.

The sigma terms for the octet states are compared with three previous lattice determi-nations [2, 37, 38] in Tab. IX. Somewhat outdated, more-phenomenological, results from [39–41] are not discussed any longer since they did neither consider volume effects nor discretization effects. A first significant chiral analysis with the consideration of volume effects in [32] is however included in that table, which work was triggered by the Lattice data of [37, 38]. Further results from our group [1, 9] are not included in the table, because they rest on partially-outdated Lattice data. While in [9] a first full analysis based on our chiral framework was documented, some ensembles with flavor-symmetric quark masses were observed to have a significant pull of the fit into a direction which is not confirmed by our improved data set. Our previous fit [9] was challenged by some ensembles, in particular D450, on which the nucleon and lambda masses were not reproduced convincingly. A 5 percent discrepancy from the lattice results was deemed troublesome, given the otherwise quite impressive reproduction of the data set. Such tensions have now been lifted with our updated data set, in particular with the updated baryon masses on the X251, X250 and X450 ensembles and the omission of the non-CLS ensemble RQCD017.

Our values for the non-strange sigma terms are in reasonable agreement with the lattice results [2, 7, 42, 43]. In particular, we obtain a rather small value for the pion-nucleon sigma term $\sigma_{\pi N} = 43.5(9)$ MeV, which is compatible with the seminal result $\sigma_{\pi N} = (45 \pm 8)$ MeV of Gasser, Leutwyler and Sainio in [44]. While our $\sigma_{\pi N}$ also agrees well with the latest lattice results from the Mainz group [7], with a value of $43.7(1.2)(3.4)$ MeV it is in tension with the current empirical value $\sigma_{\pi N} = 58(5)$ MeV from [45–48]. To further probe this, we have performed a dedicated constrained fit, in which we imposed a strict range of the sigma term to be larger than 53 MeV. The best $\chi^2$ per degree of freedom we got was about 1.25, a value considerably worse than the one of our favorite fit scenario with 0.988.

Our estimate for the strangeness sigma term of the nucleon with $\sigma_{sN} = -1(8)$ MeV is somewhat smaller than the former lattice average $\sigma_{sN} = 40(10)$ MeV obtained in [49], and also with the latest result from Mainz [7] who gave a value of $28.6(6.2)(7.0)$ MeV. For the strangeness sigma terms of the remaining octet states there appears to be an agreement with the values obtained by the Regensburg group.

There is less empirical information on the sigma terms of the decuplet baryons. Direct determination form lattice groups exist for the Omega baryon only [2]. Recent values for

|  | [40] | [41] | [50] | [32] | current Fit |
|---|---|---|---|---|---|
| $\sigma_{\pi\Delta}$ | 55(4)(18) | 34(3) | 28(1)(8) | 38(2) | 60.8(9) |
| $\sigma_{\pi\Sigma^*}$ | 39(3)(13) | 28(2) | 22(2)(9) | 27(1) | 32.7(6) |
| $\sigma_{\pi\Xi^*}$ | 22(3)(7) | 18(4) | 11(2)(6) | 14(1) | 5.5(8) |
| $\sigma_{\pi\Omega}$ | 5(2)(1) | 10(4) | 5(2)(2) | 4(2) | 15.1(4) |
| $\sigma_{s\Delta}$ | 56(24)(1) | 41(41) | 88(22)(3) | 43(27) | -16(16) |
| $\sigma_{s\Sigma^*}$ | 160(28)(7) | 211(44) | 243(24)(31) | 194(15) | 233(4) |
| $\sigma_{s\Xi^*}$ | 274(32)(9) | 373(53) | 391(24)(67) | 421(14) | 446(15) |
| $\sigma_{s\Omega}$ | 360(34)(26) | 510(50) | 528(26)(101) | 431(15) | 349(8) |

TABLE X. Pion and strangeness sigma terms of the baryon decuplet states in units of MeV. A comparison with various theoretical predictions [32, 40, 41, 50] is provided.

the omega baryon have been predicted by the Regensburg group

$$\sigma_{\pi\Omega} = 6.9^{(5.3)}_{(4.3)} \text{ MeV}, \qquad \sigma_{s\Omega} = 421^{(89)}_{(59)} \text{ MeV}, \qquad (6)$$

both values appear compatible with our results in Tab. X. The sigma terms for the baryon decuplet states are compared with three previous more-phenomenological extrapolation results [40, 41, 50]. Our values in Tab. X are consistent with the previous analysis [41], based on the same framework used in this work. We observe that even though the inclusion of finite volume effects is instrumental to determine the values of the symmetry-preserving counter-terms, the overall effect on the baryon decuplet masses at large volumes turns out to be rather small. There is also quite good consistency found with the decuplet sigma terms obtained in [40, 50].

## VI. SUMMARY AND OUTLOOK

In this work we extrapolated the baryon masses on CLS ensembles to their flavor chiral SU(3) limit. Critical to this challenge was the availability of precise values of the baryon masses along the flavor-symmetric ensembles. In a first step we generated updated and more significant results thereof.

Based on the flavor-SU(3) chiral Lagrangian with baryon octet and decuplet fields we extracted a full set of LEC that become relevant at N³LO and lead to an accurate reproduction of the dataset with a smaller residual uncertainty than we had used previously for the baryon masses of about MeV only. For the first time we obtained a chiral limit value of the pion and kaon decay constants, $f = 82.4(7)$ MeV, from such data that is compatible with what is expected from phenomenology. Such a value came along with a strange quark mass over light quark mass ratio of 27.6(1) compatible with the current FLAG value. All LEC that drive the s- and p-wave meson-baryon interactions at NLO are predicted with reasonably-small uncertainties. For the chiral-SU(3) limit masses of the baryon octet and decuplet states we predict values of 840(16) MeV and 1091(14) MeV respectively.

Despite the convincing reproduction of the Lattice data set, we now obtain a pion-nucleon sigma term, $\sigma_{\pi N} = 44(1)$ MeV that is much below the empirical value of 58(6) MeV that was extracted from experimental pion-nucleon scattering data, yet consistent with other lattice determinations of this quantity. It remains an open puzzle how such a result can be reconciled with Lattice QCD results.

**ACKNOWLEDGMENTS**

We thank John Bulava and Daniel Mohler for useful discussions. We would like to thank the CLS consortium for providing gauge configurations, in particular to Enno Scholz for helping us retrieve those used in this work.

**APPENDIX: THE CHIRAL LAGRANGIAN**

We use the chiral Lagrangian from [9] relevant in a computation of the baryon octet and decuplet masses at next-to-next-to-next-to-leading order (N³LO). The terms considered go back to a series of works [1, 32, 40, 41, 50–65]. The applied conventions follow from the

leading order terms

$$\mathcal{L}^{(1)} = \text{tr}\left\{\bar{B}\left(i\,D^\alpha\,\gamma_\alpha - M_{[8]}\right)B\right\} - \text{tr}\left\{\bar{B}_\mu \cdot \left(\left(i\,D^\alpha\gamma_\alpha - M_{[10]}\right)g^{\mu\nu}\right.\right.$$
$$\left.\left. - i\left(\gamma^\mu D^\nu + \gamma^\nu D^\mu\right) + \gamma^\mu(i\,D^\alpha\gamma_\alpha + M_{[10]})\gamma^\nu\right)B_\nu\right\}$$
$$+ F\,\text{tr}\left\{\bar{B}\,\gamma^\mu\gamma_5\,[i\,U_\mu, B]\right\} + D\,\text{tr}\left\{\bar{B}\,\gamma^\mu\gamma_5\,\{i\,U_\mu, B\}\right\}$$
$$+ C\left(\text{tr}\left\{(\bar{B}_\mu \cdot i\,U^\mu)\,B\right\} + \text{h.c.}\right) + H\,\text{tr}\left\{(\bar{B}^\mu \cdot \gamma_5\,\gamma_\nu B_\mu)\,i\,U^\nu\right\},$$

$$\Gamma_\mu = \tfrac{1}{2}u^\dagger\left(\partial_\mu u\right) + \tfrac{1}{2}u\left(\partial_\mu u^\dagger\right), \qquad U_\mu = \tfrac{1}{2}u^\dagger\left(\partial_\mu e^{i\frac{\Phi}{f}}\right)u^\dagger, \qquad u = e^{i\frac{\Phi}{2f}},$$
$$D_\mu B = \partial_\mu B + \Gamma_\mu B - B\,\Gamma_\mu, \tag{7}$$

for the baryon spin 1/2 and 3/2 fields, $B$ and $B_\mu$. The Goldstone boson sector of the Lagrangian is

$$\mathcal{L}^{(2)} = -f^2\,\text{tr}\,U_\mu U^\mu + \frac{1}{2}f^2\,\text{tr}\,\chi_+,$$
$$\mathcal{L}^{(4)} = 16\,L_1\left(\text{tr}\,U_\mu U^\mu\right)^2 + 16\,L_2\,\text{tr}\,U_\mu U_\nu\,\text{tr}\,U^\mu U^\nu + 16\,L_3\,\text{tr}\,U_\mu U^\mu U_\nu U^\nu$$
$$- 8\,L_4\,\text{tr}\,U_\mu U^\mu\,\text{tr}\,\chi_+ - 8\,L_5\,\text{tr}\,U_\mu U^\mu\,\chi_+ + 4\,L_6\left(\text{tr}\,\chi_+\right)^2$$
$$+ 4\,L_7\left(\text{tr}\,\chi_-\right)^2 + 2\,L_8\,\text{tr}\left(\chi_+\chi_+ + \chi_-\chi_-\right),$$
$$\chi_\pm = \tfrac{1}{2}\left(u\,\chi_0\,u \pm u^\dagger\,\chi_0\,u^\dagger\right), \qquad \chi_0 = 2\,B_0\,\text{diag}(m_u, m_d, m_s), \tag{8}$$

in terms of the eight LEC of Gasser and Leutwyler $L_{1-8}$. The next-to-leading terms in the baryon part of the chiral Lagrangian that contribute to the baryon masses at the one-loop level are

$$\mathcal{L}_\chi^{(2)} = 2\,b_0\,\text{tr}\left(\bar{B}\,B\right)\text{tr}\left(\chi_+\right) + 2\,b_D\,\text{tr}\left(\bar{B}\,\{\chi_+, B\}\right) + 2\,b_F\,\text{tr}\left(\bar{B}\,[\chi_+, B]\right)$$
$$- 2\,d_0\,\text{tr}\left(\bar{B}_\mu \cdot B^\mu\right)\text{tr}(\chi_+) - 2\,d_D\,\text{tr}\left((\bar{B}_\mu \cdot B^\mu)\,\chi_+\right),$$
$$\mathcal{L}_S^{(2)} = -\frac{1}{2}g_0^{(S)}\,\text{tr}\left\{\bar{B}\,B\right\}\text{tr}\left\{U_\mu U^\mu\right\} - \frac{1}{2}g_1^{(S)}\,\text{tr}\left\{\bar{B}\,U^\mu\right\}\text{tr}\left\{U_\mu B\right\}$$
$$- \frac{1}{4}g_D^{(S)}\text{tr}\left\{\bar{B}\,\{\{U_\mu, U^\mu\}, B\}\right\} - \frac{1}{4}g_F^{(S)}\text{tr}\left\{\bar{B}\,[\{U_\mu, U^\mu\}, B]\right\}$$
$$+ \frac{1}{2}h_1^{(S)}\,\text{tr}\left\{\bar{B}_\mu \cdot B^\mu\right\}\text{tr}\left\{U_\nu U^\nu\right\} + \frac{1}{2}h_2^{(S)}\,\text{tr}\left\{\bar{B}_\mu \cdot B^\nu\right\}\text{tr}\left\{U^\mu U_\nu\right\}$$
$$+ h_3^{(S)}\,\text{tr}\left\{(\bar{B}_\mu \cdot B^\mu)(U^\nu U_\nu)\right\} + \frac{1}{2}h_4^{(S)}\,\text{tr}\left\{(\bar{B}_\mu \cdot B^\nu)\{U^\mu, U_\nu\}\right\}$$
$$+ h_5^{(S)}\,\text{tr}\left\{(\bar{B}_\mu \cdot U_\nu)(U^\nu \cdot B^\mu)\right\}$$
$$+ \frac{1}{2}h_6^{(S)}\,\text{tr}\left\{(\bar{B}_\mu \cdot U^\mu)(U^\nu \cdot B_\nu) + (\bar{B}_\mu \cdot U^\nu)(U^\mu \cdot B_\nu)\right\},$$

$$\begin{aligned}
\mathcal{L}_V^{(2)} = &-\frac{1}{4} g_0^{(V)} \left( \text{tr}\left\{\bar{B} i \gamma^\mu D^\nu B\right\} \text{tr}\left\{U_\nu U_\mu\right\} \right) \\
&-\frac{1}{8} g_1^{(V)} \left( \text{tr}\left\{\bar{B} U_\mu\right\} i \gamma^\mu \text{tr}\left\{U_\nu D^\nu B\right\} + \text{tr}\left\{\bar{B} U_\nu\right\} i \gamma^\mu \text{tr}\left\{U_\mu D^\nu B\right\} \right) \\
&-\frac{1}{8} g_D^{(V)} \left( \text{tr}\left\{\bar{B} i \gamma^\mu \{\{U_\mu, U_\nu\}, D^\nu B\}\right\} \right) \\
&-\frac{1}{8} g_F^{(V)} \left( \text{tr}\left\{\bar{B} i \gamma^\mu [\{U_\mu, U_\nu\}, D^\nu B]\right\} \right) \\
&+\frac{1}{4} h_1^{(V)} \left( \text{tr}\left\{\bar{B}_\lambda \cdot i \gamma^\mu D^\nu B^\lambda\right\} \text{tr}\left\{U_\mu U_\nu\right\} \right) \\
&+\frac{1}{4} h_2^{(V)} \left( \text{tr}\left\{ \left(\bar{B}_\lambda \cdot i \gamma^\mu D^\nu B^\lambda\right) \{U_\mu, U_\nu\}\right\} \right) \\
&+\frac{1}{4} h_3^{(V)} \Big( \text{tr}\Big\{ \left(\bar{B}_\lambda \cdot U_\mu\right) i \gamma^\mu \left(U_\nu \cdot D^\nu B^\lambda\right) \\
&\qquad\qquad + \left(\bar{B}_\lambda \cdot U_\nu\right) i \gamma^\mu \left(U_\mu \cdot D^\nu B^\lambda\right) \Big\} \Big) + \text{h.c.}\,,
\end{aligned} \qquad (9)$$

where we emphasize the dual role played by the LEC in (9). At N$^3$LO two further sets of LEC are needed:

$$\begin{aligned}
\mathcal{L}_\chi^{(3)} = &\ \zeta_0 \,\text{tr}\big(\bar{B}\left(i\left(D^\alpha \gamma_\alpha\right) - M_{[8]}\right) B\big) \text{tr}(\chi_+) + \zeta_D \,\text{tr}\big(\bar{B}\left(i\left(D^\alpha \gamma_\alpha\right) - M_{[8]}\right) \{\chi_+, B\}\big) \\
&+ \zeta_F \,\text{tr}\big(\bar{B}\left(i\left(D^\alpha \gamma_\alpha\right) - M_{[8]}\right) [\chi_+, B]\big) - \xi_0 \,\text{tr}\big(\bar{B}_\mu \cdot \left(i\left(D^\alpha \gamma_\alpha\right) - M_{[10]}\right) B^\mu\big) \text{tr}(\chi_+) \\
&- \xi_D \,\text{tr}\big(\big(\bar{B}_\mu \cdot \left(i\left(D^\alpha \gamma_\alpha\right) - M_{[10]}\right) B^\mu\big) \chi_+\big), \\
\mathcal{L}_\chi^{(4)} = &\ c_0 \,\text{tr}\left(\bar{B} B\right) \text{tr}\left(\chi_+^2\right) + c_1 \,\text{tr}\left(\bar{B} \chi_+\right) \text{tr}\left(\chi_+ B\right) \\
&+ c_2 \,\text{tr}\left(\bar{B} \{\chi_+^2, B\}\right) + c_3 \,\text{tr}\left(\bar{B} [\chi_+^2, B]\right) \\
&+ c_4 \,\text{tr}\left(\bar{B} \{\chi_+, B\}\right) \text{tr}(\chi_+) + c_5 \,\text{tr}\left(\bar{B} [\chi_+, B]\right) \text{tr}(\chi_+) \\
&+ c_6 \,\text{tr}\left(\bar{B} B\right) \left(\text{tr}(\chi_+)\right)^2 \\
&- e_0 \,\text{tr}\left(\bar{B}_\mu \cdot B^\mu\right) \text{tr}\left(\chi_+^2\right) - e_1 \,\text{tr}\left(\left(\bar{B}_\mu \cdot \chi_+\right)\left(\chi_+ \cdot B^\mu\right)\right) \\
&- e_2 \,\text{tr}\left(\left(\bar{B}_\mu \cdot B^\mu\right) \chi_+^2\right) - e_3 \,\text{tr}\left(\left(\bar{B}_\mu \cdot B^\mu\right) \chi_+\right) \text{tr}(\chi_+) \\
&- e_4 \,\text{tr}\left(\bar{B}_\mu \cdot B^\mu\right) \left(\text{tr}(\chi_+)\right)^2,
\end{aligned} \qquad (10)$$

where again we focus on terms that contribute to the baryon masses.

|  | Fit [9] | current Fit |  | Fit [9] | current Fit |
|---|---|---|---|---|---|
| $c_0$ [GeV$^{-3}$] | 0.3884(8) | 0.4506(241) | $c_1$ [GeV$^{-3}$] | 0.2201(7) | 0.2869(55) |
| $c_2$ [GeV$^{-3}$] | -0.4521(3) | -0.5175(102) | $c_3$ [GeV$^{-3}$] | 0.4591(6) | 0.4755(25) |
| $c_4$ [GeV$^{-3}$] | 0.3281(3) | 0.1779(114) | $c_5$ [GeV$^{-3}$] | -0.3176(3) | -0.2128(260) |
| $c_6$ [GeV$^{-3}$] | -0.2893(5) | -0.3985(280) | $\zeta_0$ [GeV$^{-2}$] | 0.3823(241) | 0.3344(113) |
| $\zeta_D$ [GeV$^{-2}$] | -0.1145(5) | 0.0236(79) | $\zeta_F$ [GeV$^{-2}$] | 0.0504(33) | 0.0615(5) |
| $e_0$ [GeV$^{-3}$] | -0.2560(9) | -0.2024(449) | $e_1$ [GeV$^{-3}$] | 0.3806(75) | 0.5418(628) |
| $e_2$ [GeV$^{-3}$] | 0.3005(83) | 0.1927(443) | $e_3$ [GeV$^{-3}$] | 0.0314(1) | -0.1044(998) |
| $e_4$ [GeV$^{-3}$] | 0.0897(6) | -0.3479(427) | $\mu$ [GeV] | 0.701(2) | 0.9102(321) |
| $\xi_0$ [GeV$^{-2}$] | 0.2174(11) | 0.2964(340) | $\xi_D$ [GeV$^{-2}$] | -0.1923(141) | 0.2551(74) |

TABLE XI. LEC from a fit to baryon masses on CLS ensembles. Not all LEC in this table are independent. There are 6+3+2 independent parameters only [1, 4]. The large-$N_c$ sum rules are imposed on the LEC at a given renormalization scale $\mu$. The optimal value for $\mu$, at which the sum rules work best is determined in the global fit.